\documentclass[11pt]{article}

\pdfoutput=1

\usepackage{graphics, color,soul}
\usepackage{graphicx}
\usepackage{amssymb}

\usepackage{mathtools}

\usepackage{booktabs}
\usepackage[english]{babel}
\usepackage{amsmath,amssymb,amsbsy,amstext, amsthm, simplewick}
\usepackage{hyperref}
\usepackage{tikz}

\usetikzlibrary{decorations.pathmorphing,shapes.misc}
\tikzset{snake it/.style={decorate, decoration=snake}}
\tikzset{cross/.style={cross out, draw=black, minimum size=2*(#1-\pgflinewidth), inner sep=0pt, outer sep=0pt},
cross/.default={1pt}}

\usepackage{amsfonts}
\usepackage{amssymb}
\usepackage{upgreek}
\usepackage{simplewick}
 \usepackage{exscale,relsize}
\usepackage{mathtools}
\usepackage{comment}

\usepackage[margin=1cm,labelfont={sf,bf,scriptsize},textfont={sf,scriptsize}]{caption}

\usepackage{colortbl}
\definecolor{lightgreen}{cmyk}{0.2, 0, 0.2, 0.2}
\definecolor{lightgray}{cmyk}{0.1,0.2,0,0.1}
\definecolor{lightgray2}{cmyk}{0.1,0.1,0,0.1}

\makeatletter
\newlength{\apb@width}
\newcommand{\autoparbox}[2][c]{\settowidth{\apb@width}{#2}\parbox[#1]{\apb@width}{#2}}

\makeatother

\numberwithin{equation}{section}

\def\beq{\begin{equation}}
\def\eeq{\end{equation}}

\def\bea{\begin{eqnarray}}
\def\eea{\end{eqnarray}}

\def\beq{\begin{equation}}
\def\eeq{\end{equation}}
\def\be{\begin{equation}}
\def\ee{\end{equation}}
\def\bea{\begin{eqnarray}}
\def\eea{\end{eqnarray}}

\def\0{{\vec{0}}}

\DeclareRobustCommand{\SkipTocEntry}[4]{}

\def\beq{\begin{equation}}
\def\eeq{\end{equation}}

\def\ba#1\ea{\begin{align}#1\end{align}}
\def\bg#1\eg{\begin{gather}#1\end{gather}}
\newcommand{\bseq}{\begin{subequations}}
\newcommand{\eseq}{\end{subequations}}

\renewcommand{\t}{\tilde}
\newcommand{\ap}{{\alpha '}}

\DeclareSymbolFont{extraup}{U}{zavm}{m}{n}
\DeclareMathSymbol{\varheart}{\mathalpha}{extraup}{86}
\DeclareMathSymbol{\vardiamond}{\mathalpha}{extraup}{87}

\def\({\left(}
\def\){\right)}
\def\[{\left[}
\def\]{\right]}

\begin{document}

\begin{titlepage}

\setcounter{page}{1} \baselineskip=15.5pt \thispagestyle{empty}

\vbox{\baselineskip14pt
}
{~~~~~~~~~~~~~~~~~~~~~~~~~~~~~~~~~~~~
~~~~~~~~~~~~~~~~~~~~~~~~~~~~~~~~~~
~~~~~~~~~~~ }

\bigskip\

\vspace{2cm}
\begin{center}
{\fontsize{19}{36}\selectfont  \sc
Varying dilaton as a tracer of classical string interactions
}
\end{center}

\vspace{0.6cm}

\begin{center}
{\fontsize{13}{30}\selectfont  Matthew Dodelson$^{1},$ Eva Silverstein$^{1,2},$ and Gonzalo Torroba$^3$}
\end{center}


\begin{center}
\vskip 8pt

\textsl{
\emph{$^1$Stanford Institute for Theoretical Physics, Stanford University, Stanford, CA 94306}}



\vskip 7pt
\textsl{ \emph{$^2$Kavli Institute for Particle Astrophysics and Cosmology, Stanford, CA 94025}}

\vskip 7pt
\textsl{ \emph{$^3$Centro At\'omico Bariloche and CONICET, Bariloche, R8402AGP, Argentina}}

\end{center}

\vspace{0.5cm}
\hrule \vspace{0.1cm}
{ \noindent \textbf{Abstract}  We analyze tree-level string amplitudes in a linear dilaton background, motivated by its use as a gauge-invariant tracer of string interactions in scattering experiments and  its genericity among simple perturbative string theory limits.  A simple case is given by a lightlike dependence for the dilaton.  The zero mode of the embedding coordinate in the  direction of dilaton variation requires special care.  Employing Gaussian wave packets and a well-defined modification of the dilaton profile far from the dominant interaction region, we obtain finite results which explicitly reproduce the interaction timescales expected from joining and splitting interactions involving oscillating strings in simple string scattering processes.   There is an interesting interplay between the effects of the linear dilaton and the $i\epsilon$ prescription. In more general circumstances this provides a method for tracing the degree of non-locality in string interactions, and it gives a basis for further studies of perturbative supercritical string theory at higher loop order.  
\vspace{0.3cm}
 \hrule

\vspace{0.6cm}}
\end{titlepage}

\tableofcontents

\section{Introduction and setup}

Perturbative string theory is a mature subject, but essential features remain under development.  
These include its implications for horizon physics, in black hole backgrounds and cosmology.  

It is useful to develop tools applicable to these issues.
In this note, we analyze tree-level string perturbation theory in a linear dilaton background.  This case is simple enough to be tractable, while more generic than string theory in Minkowski spacetime.  It has various applications, including some relevant in horizon physics such as \cite{sixpoints, SCloops}.  For example, for a generic total dimensionality $D>D_{\text{crit}}$, the simplest classical solution of the theory is a runaway to weak coupling along a timelike linear dilaton trajectory, corresponding to the rolling scalar solution sourced by a positive scalar potential.  As another application, in black hole physics it is crucial to understand the degree of non-locality of interactions; a dilaton background can serve as a tracer of the spacetime location of interactions.  

The computation of tree-level string scattering in a nontrivial dilaton background has some technical subtleties having to do with the zero mode of the embedding coordinate along the linear dilaton.  This problem was analyzed in \cite{LDprevious}, but we differ from those authors on the appropriate treatment of the zero mode.  Numerous works have treated spacelike linear dilaton backgrounds in various contexts.\footnote{See e.g. \cite{otherLDscatt}\ and references therein for some examples.}   

\subsection{Worldsheet path integral}

Let us begin by setting up the computation of scattering in a linear dilaton background.

\subsection{Background and vertex operators}

The bosonic linear dilaton CFT has Euclidean worldsheet action
\be\label{wsac}
S=\frac{1}{4\pi\ap} \int d^2 \sigma \sqrt{g} \left(g^{ab} \partial_a X^\mu \partial_b X_\mu+\ap R\, \Phi(X) \right)\,,
\ee
where
\be
\Phi(X)= V_\mu X^\mu\,,
\ee
and $R$ is the worldsheet scalar curvature.\footnote{We follow the conventions in \cite{JoeBook}; the worldsheet is euclidean, and the $D$-dimensional target space $X^\mu$ has signature $(-+\ldots +)$.}  The string coupling 
\be\label{gsdef}
g_s=g_0 e^{V_\mu X^\mu}
\ee
varies in spacetime.    It satisfies
\be\label{VD}
V_\mu V^\mu = \frac{26-D}{6\ap}\,.
\ee
in order to cancel the Weyl anomaly, giving total central charge zero.  
This coincides with the equation of motion for the dilaton derived from the two-derivative approximation to the spacetime effective action, but it is an exact classical solution to string theory regardless of the size of $V$.

In the critical dimension $D=26$, the linear dilaton solution is lightlike, varying say along the direction $X^+=(T+X)/\sqrt{2}$ for some spatial direction $X$.  This has a tunable parameter $V^-$.  In this lightlike case, any function $g_s(X^+)$ purely depending on $X^+$ solves the field equations.  We will focus on this case.     

We would like to analyze the worldsheet path integral
\be\label{wsPI}
\int \frac{Dg\, DX}{V_{\text{diff}}\times V_{\text{Weyl}}} \prod_I\int d^2\sigma_I \sqrt{g(\sigma_I)} {\cal V}_I [X(\sigma_I)] e^{- S}
\ee
where the integrated vertex operators inserted here are Weyl and diffeomorphism invariant.   

As we will review shortly, these operators are proportional to the varying string coupling (\ref{gsdef}).   As a result of this and the curvature term in the worldsheet action (\ref{wsac}), the path integrand varies exponentially with the embedding coordinate in the linear dilaton direction.  Nonzero modes are suppressed by the Gaussian factor in $e^{-S}$ (\ref{wsac}).  But the zero mode integral in this direction is more divergent than it is for $V=0$ (where it gives the energy-momentum conserving delta function).   We will address this with two additional ingredients.  First, we will work with nontrivial wavepackets aiming the strings to scatter in a limited interaction region.  This does not fully control the zero mode integral, however: taking into account the positivity of the energy of incoming and outgoing strings, the wavepackets have nontrivial position-space tails (albeit highly suppressed) in the strong coupling region.  Because of this, we we will also modify the dilaton profile so as to smoothly asymptote to a constant value, small enough to maintain the strong suppression of the wavepacket tails.

Let us focus on the CFT on the complex plane $(z, \bar z)$, pushing the curvature to infinity. The scalar two-point function is the same as in the $V_\mu=0$ case
\be
\langle X^\mu(z, \bar z) X^\nu(0,0)\rangle=-\frac{\ap}{2} \eta^{\mu\nu}\,\log |z|^2\,.
\ee
Normal ordering is defined as usual by subtracting Wick contractions,
\be\label{eq:normal}
: X^\mu(z_1, \bar z_1) X^\nu(z_2, \bar z_2):=X^\mu(z_1, \bar z_1) X^\nu(z_2, \bar z_2)+\frac{\ap}{2} \eta^{\mu\nu}\,\log |z_{12}|^2\,,
\ee
and so on.

The holomorphic and antiholomorphic parts of the stress tensor are
\bea\label{eq:Tz}
T(z) &=&-\frac{1}{\ap} :\partial X^\mu \partial X_\mu:+ V_\mu \partial^2 X^\mu\,, \nonumber\\
\tilde T(\overline{z}) &=&-\frac{1}{\ap} :\bar \partial X^\mu \bar \partial X_\mu:+ V_\mu \bar \partial^2 X^\mu\,.
\eea
Let us compute some basic OPEs, using (\ref{eq:normal}):
\bea\label{eq:OPE}
T(z) X^\mu(0,0)& =&\frac{\ap}{2}\,\frac{V^\mu}{z^2}+\frac{\partial X^\mu(0)}{z}+\ldots \nonumber\\
T(z) \partial X^\mu(0)& =& \ap \frac{V^\mu}{z^3}+\frac{\partial X^\mu(0)}{z^2}+\ldots \\
T(z) : e^{i p \cdot X(0)}:&=&\frac{\ap}{4} (p_\mu p^\mu+2 i V_\mu p^\mu)\frac{ : e^{i p \cdot X(0)}:}{z^2}+\ldots \nonumber
\eea
The $TT$ correlator gives
\be
T(z) T(0) = (D+6\ap V_\mu V^\mu )\frac{1}{2z^4}+\ldots\,,
\ee
so that the central charge is $c=D+6\ap V_\mu V^\mu$. The antiholomorphic sector has $\tilde c= c$.

The linear dilaton introduces nontensor terms in the OPE of $X^\mu$ and also of $\partial X^\mu$, and it shifts the conformal weight of the exponential vertex operator $:e^{ip\cdot X}:$ to
\be\label{eq:hexp}
h= \tilde h=\frac{\ap}{4} (p_\mu p^\mu+2 i V_\mu p^\mu)\,.
\ee
Imposing that this tachyon vertex operator be dimension (1,1) implies
\be\label{onshell}
p_\mu p^\mu+2 i V_\mu p^\mu=\frac{4}{\ap},
\ee
For more general vertex operators we find
\be\label{massshell}
p_\mu p^\mu+2 i V_\mu p^\mu=-M^2,
\ee
where $M^2=\frac{4}{\ap}(\hat h-1)$, with $\hat h$ the dimension of the vertex operator at $p=0$.     

As discussed in the Appendix, the $V$-dependent terms in (\ref{eq:OPE}) implies that the momentum changes under a conformal transformation. Therefore, in the following we will need to keep track of which frame we are working. We denote the momentum in the complex plane (or the sphere with curvature pushed to infinity) by $p_\mu$. The momentum in the cylinder frame, with coordinate $w= i \log z$, then becomes (see (\ref{eq:pcyl}))
\be\label{hatpdef}
\hat p_\mu=p_\mu+i V_\mu\,.
\ee
The on-shell condition simplifies to
\be\label{hatpdispersion}
\hat p^2=-M^2-V^2
\ee
For a lightlike linear dilaton, this is simply $\hat p^2=-M^2$, that is $\hat p$ satisfies the on-shell condition in the absence of the linear dilaton.  In terms of this decomposition, we can write the vertex operator for a momentum eigenstate as
\be\label{Vform}
{\cal V}=g_0\,: e^{V\cdot X} e^{i\hat p\cdot X}{\cal O}_{h, \tilde h}:
\ee
where $\hat p$ satisfies (\ref{hatpdispersion}) and ${\cal O}$ is an operator of the specified dimensions (satisfying additional physical state conditions determined by its OPE with the stress-energy tensor).  The vertex operator thus decomposes into a factor of $g_s(X)$ times an operator similar to the corresponding Minkowski spacetime vertex operator.  For example for the tachyon vertex operator, ${\cal O}$ is trivial and in a lightlike linear dilaton background $\hat p$ satisfies the dispersion relation for the tachyon vertex operator in Minkowski spacetime.

It is useful to keep in mind the spacetime description of the linear dilaton background, where these results arise very simply. In string frame, the action for a scalar field of mass $M$ is
\be\label{targetEFT}
S= -\frac{1}{2}\int d^D x\, \sqrt{g}\, e^{-2\Phi}\left((\partial \varphi)^2+M^2\varphi^2 \right)\,.
\ee
For a plane wave $\varphi(x) \sim e^{i p\cdot x}$, the on-shell condition is (\ref{massshell}).  As with the dilaton background itself, this low energy EFT description does not make clear the exactness of the mass shell condition in the linear dilaton background, which follows from the full worldsheet CFT analysis just reviewed.  

\subsection{Light cone gauge wavefunctions in the linear dilaton background}\label{sec:lc}

One interesting application of a varying dilaton is to help diagnose the level of nonlocality in string interactions \cite{sixpoints,lennyspreading, grossmende, BHpaper}.  
In this section we determine the effect of the linear dilaton on string spreading in light cone gauge \cite{lennyspreading}.  
We start by defining two longitudinal directions $X^\pm$ and $D-2$ transverse directions $X_\perp$.  
We fix the gauge
\be\label{Xminus}
X^-=x^-+\ap \hat p^-\tau 
\ee
and impose the constraints (cf. (\ref{eq:Tz}))
\be\label{constraint}
2(\partial_\tau\pm\partial_\sigma) X^+(\partial_\tau\pm \partial_\sigma) X^-=2\alpha' \hat p^-(\partial_\tau\pm \partial_\sigma)X^+=-\ap V\cdot (\partial_\tau\pm \partial_\sigma)^2 X +((\partial_\tau\pm \partial_\sigma) X_\perp)^2.
\ee
Writing the mode expansion
\be\label{modeexp}
X^\mu = x^\mu+\ap \hat p^\mu\tau +i\sqrt{\frac{\ap}{2}}\sum_{n\neq 0} \left(\frac{\alpha^\mu_n}{n}e^{-in(\tau+\sigma)}+\frac{\t \alpha^\mu_n}{n}e^{-in(\tau-\sigma)}\right)
\ee
and rearranging (\ref{constraint}) in the form
\begin{align}
\alpha' (\hat{p}^--V^-\partial_\tau)\partial_\sigma X^+=-\alpha'V_\perp\cdot \partial_\tau\partial_\sigma X_\perp+\partial_\tau X_\perp\cdot \partial_\sigma X_\perp,
\end{align}
we can solve for the longitudinal oscillators:
\be\label{Valpha}
\alpha_n^+=\frac{\sqrt{\frac{2}{\ap}} L_n^\perp+in V_\perp\cdot \alpha_{\perp n} }{\hat p^-+in V^-},\hspace{10 mm}n\not=0.
\ee
where $L_n^\perp =\frac{1}{2} \sum_m \alpha^i_m \alpha^i_{n-m}$ is the Virasoro generator for the transverse sector in the unperturbed $V=0$ theory.  A similar expression applies for the tilded oscillators.
Finally, a slight generalization of the calculation in \cite{lennyspreading} gives the expectation value of the variance of the embedding
of the string in the $X^+$ direction:
\begin{align}\label{Vvariance}
\langle \hat{p}^-| (X^+-x^+)^2|\hat{p}^-\rangle &=- \frac{\ap}{2} \sum_{n,-m>0}\frac{1}{nm}\langle \hat{p}^-|[\alpha^+_n,\alpha^+_m]|\hat{p}^-\rangle e^{-i(n+m)\sigma}+(\alpha\to \tilde{\alpha})
\nonumber\\
&=-\frac{\alpha'}{2}\sum_{n,-m>0} \frac{\langle \hat{p}^-|\frac{2}{\ap n m}[L^\perp_n, L^\perp_m]-V_\perp^2[\alpha^i_n, \alpha^i_m] |\hat{p}^-\rangle e^{-i(n+m)\sigma}}{(\hat p^-+i n V^-)(\hat p^-+i m V^-)}+(\alpha\to \tilde{\alpha}).
\end{align}
Next, applying
\be\label{Vircentral}
[{L_n^\perp}, {L_m^\perp}]=\dots+\frac{D-2}{12}n^3\delta_{n+m}, ~~~[{\alpha_n^i}, {\alpha_m^i}]=n{\delta_{n+m}}
\ee
we obtain 
\be\label{DelXV}
\langle\hat{p}^-|(\Delta X^+)^2|\hat{p}^-\rangle \sim \sum_{n>0} \frac{(2+V^+V^-\ap) n}{{{\hat p}^{-2}}+n^2{V^-}^2}\,,
\ee
where we used (\ref{VD}) to simplify the numerator.  Below, we will mostly focus on the case with $V^+=0, V^-\ne 0$.  

The divergence in this sum is operationally cut off by the time resolution of whatever detects this effect \cite{lennyspreading, BHpaper}.
The expression (\ref{DelXV}) indicates that the $V^-$ component of the linear dilaton shuts off the quadratic divergence in the mode sum, at $n\sim \hat{p}^-/V^-$.  Conversely, for a detector corresponding to a maximum mode number $n_{\text{max}}\ll \hat{p}^-/V^-$, the spreading prediction is well approximated by the original one in the absence of the linear dilaton.  
The upshot of this is that a weakly varying dilaton can be used as a tracer of longitudinal string spreading without significantly affecting its predicted range.   

\section{Scattering amplitudes with varying dilaton}

Let us introduce wavepackets for the strings since our goal is to use the dilaton profile to help track the geometry of string scattering.  This will also help control the zero mode integral discussed below (\ref{wsPI}).    We may scatter any linear combination of on-shell vertex operators for the strings, maintaining positive frequency to ensure that the string is either ingoing or outgoing.  
Let us start with one of the strings, labeling it string $A$, and work with massless strings (or with appropriate adjustments, any fixed-mass strings in the highly relativistic limit).  The wavepacket is taken to be Gaussian,    
\be\label{wavepacketAGauss}
\Psi_{X_{A0}^+, p_{A0}^-}(\hat p_A^-)=\frac{1}{\pi^{1/4}(\sigma^-)^{1/2}}\exp (i \hat p_A^- X_{A0}^+) \exp({-(\hat p_A^--\hat p_{A0}^-)^2/2{\sigma^-}^2})\theta(\hat p_A^-)\,.
\ee
Here $X_{A0}^+, \hat p_{A0}^-$ are parameters which describe (to good approximation) the peak position and momentum of the wavepacket.   We specify $\hat p_{A0}^-\gg \sigma^-$.  

Fourier transforming this to position space (integrating it against $e^{-i \hat p_A^- X_A^+(z_A, \bar z_A)}$) gives the wavefunction in a position basis.  This is an error function, which is approximately given by  
\begin{align}\label{Gaussplustail}
\Psi_{X_{A0}^+, \hat p_{A0}^-}( X_A^+(z_A, \bar z_A))&\simeq (\sigma^-)^{1/2} e^{i\hat p^-_{A0}(X^+_{A0}-\hat X^+(z_A, \bar z_A))}e^{-(X_A^+(z_A, \bar z_A)-X_{A0}^+)^2{\sigma^-}^2/2}\nonumber\\
&\hspace{5 mm}+(\sigma^-)^{1/2}\frac{e^{-(\hat p_{A0}^-)^2/2{\sigma^-}^2}}{i(X_A^+(z_A,\bar z_A)-X_{A0}^+){\sigma^-}+{\hat p_{A0}^-}/\sigma^-}\,.
\end{align}
The Gaussian suppression in the first term here is enough to render finite the integral over the zero mode $X^+_0$ of the embedding coordinate $X^+(z, \bar z)$ even in the presence of the exponentially growing factors arising from the linear dilaton.   

The second term in (\ref{Gaussplustail}) has an exponentially small coefficient, but a weak power law suppression in position.  If the dilaton remains linear for all $X^+$, this term in the wavepacket could access a region of arbitrarily strong coupling.   

To avoid such contributions, we can modify the background dilaton profile so that the dilaton $\Phi(X^+)$ smoothly approaches a constant well outside the interaction region supported by the Gaussian components of the wavefunctions.  That is, we keep $\Phi$ linear throughout this interaction region but let it asymptote to a constant at a very large value of $X^+$.  This constant value may easily be chosen small enough that the power law tail term in (\ref{Gaussplustail}) is highly suppressed everywhere.  Given this, we may focus on the Gaussian piece of the wavepackets supported well within the linear dilaton region of the background.\footnote{Another approach to this problem would be to introduce an exponentially growing background for a different field which becomes strong at some large value of $X^+$, analogous to a tachyon wall in Liouville theory.}  

For the Gaussian term in (\ref{Gaussplustail}), we may treat $\Phi(X^+)$ as purely linear, since as just explained the Gaussian wavepackets strongly suppress contributions from the modified region.   Let us perform the integrals in     
(\ref{wsPI}) in the following order.  First we integrate over $\tilde p_A$ as above, obtaining (\ref{Gaussplustail}).  Next, we integrate over the zero mode $x^+_0$ of the string embedding coordinate 
\be\label{defXmodes}
X^+(z, \bar z)\equiv x^+_0+\hat X^+(z, \bar z), 
\ee
and without loss of generality, center string $A$ at the origin
\be\label{Aorigin}
X^+_{A0}=0.
\ee
This gives
\bea\label{zint}
I_0 &=&\int dx^+_0 \, e^{-(x^+_0+\hat X^+(z_A, \bar z_A))^2{\sigma^-}^2/2} e^{-i x^+_0\sum_{I=1}^N(\hat p_I^--iV^-)} e^{2 V^-x^+_0} \\ 
&=&\frac{\sqrt{2\pi}}{\sigma^-} \exp\left(-\frac{1}{2{\sigma^-}^2}(\sum_{I=1}^N\hat p_I^--(N-2)i V^-)^2\right) \exp\left(i \hat X^+(z_A, \bar z_A)\left(\sum_{I=1}^N\hat p^-_I-(N-2)i V^-\right)\right)\nonumber
\eea     
The final factor in the integrand here is from the curvature term in (\ref{wsac}).   

Also doing the zero mode integrals in the directions transverse to the linear dilaton, the amplitude reduces to wavepackets $\Psi(\hat p_{I\ne A})$  for the other strings convolved with
\bea\label{PInext}
{\cal A}_{\hat p_I}&\approx& \frac{g_0^{N-2}}{(\sigma^-)^{1/2}}\delta(\sum_I \hat p_I^+)\delta^{D-2}(\sum_I \hat p_{I\perp})\exp\left(-\frac{1}{2{\sigma^-}^2}(\sum_{I=1}^N\hat p_I^--(N-2)i V^-)^2\right)\left\langle \prod_{I=N-2}^N c(z_{I})\tilde c(\bar z_{I})\right\rangle\nonumber\\
&\times&\int \prod_{I=1}^{N-3} d^2z_I\int D\hat X \, e^{-S_0} e^{2 V^- \hat X^+}|_{z_{N+1}\to\infty} \left\{e^{i \hat X^+(z_A, \bar z_A)([\sum_{I\ne A}\hat p_I^-]-(N-3)i V^-)}{\cal O}_A
\prod_{J\ne A}e^{ip_J\cdot  \hat X(z_J, \bar z_J) } {\cal O}_J\right\} \nonumber\\
\eea
where $S_0$ is the quadratic term in the worldsheet action (\ref{wsac}).  Here $z_{N+1}\to\infty$ is the point at which we have concentrated the worldsheet curvature, pushed to the point at infinity.  (If we keep it at a finite point $z_{N+1}$, the dependence on this position drops out as can be seen by consistently keeping all of the dependence on the Weyl factor $e^\omega$, $\omega=-\log|z-z_{N+1}|^2$ in the worldsheet metric.)  

This expression has several interesting properties.      
The would-be delta function in the $ p^-$ direction is replaced by a Gaussian peaked at the momentum-conserving value for $\hat p_I^-$.  The remaining path integral over the nonzero modes has effective momenta which sum to zero (including the background charge factor).  
Let us define
\be\label{hatkA}
\hat k_A=-\sum_{I\ne A} \hat p_I\,,
\ee
which would be the momentum of $A$ in the $V\to 0$ limit with energy-momentum conservation. 
Then the momenta appearing in the nonzero mode path integral are:
\bea\label{fullkin}
k_A&=& \hat k_A+(N-3)i V,  \nonumber \\
k_{J\ne A} &=&p_{J\ne A}=\hat p_J-iV, \nonumber\\
k_{N+1} &=& 2i V \,.
\eea
The Gaussian path integral over the nonzero modes $\hat X$ can now be performed in the standard way, with
the momenta (\ref{fullkin}) playing the role of sources.  For example for scattering of $N$ massless strings with
polarization vectors $ s_{\mu_I\nu_I}^{(I)}$ we obtain, up to order one constants, \cite{JoeBook}
\bea\label{Npoint}  
\mathcal A&\approx &  \frac{g_0^{N-2}}{(\sigma^-)^{1/2}}\prod_{I=1}^N s_{\mu_I\nu_I}^{(I)}\delta(\sum_I \hat p_I^+)\delta^{D-2}(\sum_I \hat p_{I\perp})\exp(-(\sum_{I=1}^N\hat p^-_I-(N-2)i V^-)^2/2{\sigma^-}^2)\\
 &\times&\left\langle \prod_{I=N-2}^N c(z_{I})\tilde c(\bar z_{I})\right\rangle \int_{\cal C} \prod_{I=1}^{N-3} d^2z_I \prod_{I<J} |z_{IJ}|^{\ap k_I\cdot k_J}\left\langle \prod_{J=1}^N[v^{\mu_J}+q^{\mu_J}(z_J)][\tilde v^{\nu_J}+\tilde q^{\nu_J}(\bar z_J)] \right\rangle\,.\nonumber
\eea
Here we followed \cite{JoeBook}, packaging the result as
\be\label{vqdef}
v^{\mu_J}(z_J)=-i\frac{\ap}{2}\sum_{I\ne J} \frac{k_I^{(\mu_I)}}{z_J-z_I}, ~~~~ q^{\mu_J}(z_J)=\partial X(z_J)-v\,,
\ee
where the $q^\mu$ correlators are given by all contractions of $\partial X$ that would arise without the exponentials.  
The subscript ${\cal C}$ denotes a contour of integration for the vertex operator positions which we will need to determine along the lines of \cite{Edepsilon}.  

The background charge at $z_{N+1}\to\infty$ drops out of this since $k_{N+1}\cdot (\sum_J k_J+k_{N+1})=0$.    
From (\ref{hatkA}-\ref{fullkin}), the kinematic invariants $k_I\cdot k_J$ are given as follows
\bea\label{KIJs}
k_I\cdot k_J &=& \hat p_I\cdot \hat p_J-iV\cdot(\hat p_I+\hat p_J), ~~~~ I, J\ne A \nonumber \\
k_I\cdot k_A &=& \hat k_A\cdot \hat p_I-i V\cdot (\hat k_A+\hat p_I)+(N-2)i V\cdot\hat p_I ,  ~~~~ I\ne A\nonumber\\
k_A^2 &=& (\hat k_A+i (N-3) V)^2 \,.
\eea    
If we take $\hat k_A$ purely lightlike, with its only nonzero component $\hat k_A^-$, then for lightlike $V^-$, we have $\hat k_A\cdot V=0$.  Using this and (\ref{hatkA}) we find $k_A^2=0$, as well as the simplification
\be\label{sumsimp}
\sum_{I\ne A} k_A\cdot k_I=-\hat k_A^2=0.
\ee
This will lead to amplitudes with complex Mandelstam invariants, shifted by $V$-dependent terms.  In the next section, we will analyze this in more detail at four points and apply it to trace the positions of splitting and joining interactions.

\section{String oscillations and splitting/joining interactions}

Let us analyze this in more detail for $2\to 2$ scattering $A+B\to \hat A+\hat B$.  As an illustration of the role of the linear dilaton in tracking string interactions, we will study a regime with the incoming strings $A$ and $B$ joining to produce  macroscopic $s$-channel strings.  These can oscillate any number of times before splitting into $\hat A$ and $\hat B$ \cite{sst, Smatrixpaper}.  This splitting interaction occurs at regular intervals in time (and hence in $X^+$), predicting corresponding values for the dilaton as it varies with $X^+$.  We will see that this prediction is precisely satisfied as a result of the $V$ dependence in the amplitude.   

In order to proceed, we need to determine the contour ${\cal C}$ for the vertex operator integrations, as in the recent treatment \cite{Edepsilon}.  
For open strings, we obtain a general formula analogous to (\ref{Npoint}), with the open string coupling $g_O=g_s^{1/2}$ substituted for the closed string coupling $g_s$.  On the disc, the Euler character is also halved, so overall the result is similar to the above closed string ampitude but with $V$ divided by 2.  Let us denote this $V_O=V/2$.  At four points, for example, one ordering ($AB\hat B \hat A$) gives a vertex operator integral
\be\label{yint}
\int_{{\cal C}_{01}} dy \,y^{2\ap k_A\cdot k_B}(1-y)^{2 \ap k_A\cdot k_{\hat A}}\,,
\ee
with ${\cal C}_{01}$ a contour from $y=0$ to $y=1$.  
In our kinematics (with $\hat p_A^+=0=\hat p_{A,\perp}$), we have
\bea\label{ksfour}
k_A\cdot k_B&=&\hat k_A\cdot \hat p_B-i V_O^- \hat p_B^+=k_{\hat A}\cdot k_{\hat B}\nonumber \\
k_A\cdot k_{\hat A} &=& \hat k_A\cdot \hat p_{\hat A}-i V_O^-\hat p_{\hat A}^+
\eea
Let us work in the center of mass frame with energies $E$ and transverse momenta $\pm q$ for the outgoers, with $q\ll E$.  Taking for simplicity the ultrarelativistic limit $E \gg 1/\sqrt{\ap}$,\footnote{Similar results are valid at smaller energies, with slightly more involved formulas.} the momenta are then approximated by
\bea\label{ksEq}
\hat k_A &=& (E+q^2/2E, -E-q^2/2E, 0, \dots, 0) \nonumber\\
\hat p_B &=& (E+q^2/2E, E+q^2/2E, 0, \dots, 0) \nonumber \\
\hat p_{\hat A} &=& (-E-q^2/2E, E, q, 0, \dots, 0) \nonumber\\
\hat p_{\hat B} &=& (-E-q^2/2E, -E, -q, 0, \dots, 0)\,,
\eea
which leads to
\bea\label{KEq}
k_A\cdot k_B&=&k_{\hat A}\cdot k_{\hat B}\simeq  -2 E^2-\sqrt{2}i V_O^- E\nonumber \\
k_A\cdot k_{\hat A} &=& \frac{q^2}{2} +i V_O^- \frac{q^2}{2\sqrt{2} E}\,.
\eea

The contour introduced in \cite{Edepsilon}\ introduces worldsheet Lorentzian time evolution near the endpoints of the $y$ integral where the spacetime propagator emerges.  The contour integral behaves near the $y=0$ endpoint as
\be\label{tauL}
 y_0^{2\alpha'k_A\cdot k_B+1}\int_0^\infty d\tau_L\,  e^{-i \tau_L (2\alpha'k_A\cdot k_B+1-i\epsilon)}, ~~~~ V_O^- \hat p_B^+=\sqrt{2} V_O^- E\ge 0.
\ee
with $y_0\ll 1$.  
At $V=0$, this converges for any real value of $k_A\cdot k_B$.  When we turn on $V$, this converges only if the imaginary part of $k_A\cdot k_B$ has the same sign as the $i\epsilon$ term, i.e. for $V_O^-\hat p_B^+=\sqrt{2} V_O^- E\ge 0 \Rightarrow V^->0$. Given this, it converges
for any value of $\hat p_B\cdot \hat k_A$ (i.e. any value of the real part of the Mandelstam $s$ variable).   
Near the $y=1$ endpoint, the contour in \cite{Edepsilon}\ runs in the opposite direction around the point $y=1$ (clockwise rather than counterclockwise) .  It can be written as
\be\label{tauLone}
(1-y_1)^{2\alpha' k_A\cdot k_{\hat{A}}+1}\int_{0}^\infty d\tau_{L1} \, e^{i\tau_{L1} (2\alpha'k_A\cdot k_{\hat A} +1+i\epsilon)} \,.
\ee
This also converges for $V^->0$ in the physical kinematics above.

If $V^-<0$, this contour prescription does not converge, but it would converge to reverse the orientation of the $\tau_L$ and $\tau_{L1}$ portions of the contour.  This corresponds to evolving in the time-reversed direction.  We will comment on this further below after unpacking some of the physics of this amplitude in the $V^->0$ dilaton background.  

For the above choice of contour combined with sign of $V^-$, the result is given by the Veneziano amplitude with the $V$-dependent shift playing the role of the $i\epsilon$ (in fact obviating the need for the latter): 
\bea\label{Aopenfour}
\mathcal A& \sim & \frac{g_{O,0}^{N-2}}{(\sigma^-)^{1/2}}\delta(\sum_I \hat p_I^+)\delta^{D-2}(\sum_I \hat p_{I\perp})\exp\left(-(\sum_{I=1}^N\hat p^-_I-(N-2)i V_O^-)^2/2{\sigma^-}^2\right)\nonumber\\
& \times &\frac{\Gamma(2{\ap} k_A\cdot k_{\hat A}+1)\Gamma(2{\ap} k_A\cdot k_{B}+1)}{\Gamma(2{\ap}k_A\cdot k_{\hat A}+2\ap k_A\cdot k_B+2)} \,.
\eea

In order to understand better the analytic structure of this result, let us expand near one of the $s$-channel poles,
\be\label{eq:poles}
\Gamma(2\ap k_A \cdot k_B+1) \propto \frac{1}{2\ap k_A \cdot k_B+n}=-\frac{1}{\ap} \frac{1}{(2E)^2-\frac{n}{\ap}+2\sqrt{2}iV_O^- E}\,,
\ee
with $n$ a positive integer. The poles correspond to a particle of mass squared $n/ \ap$; they are both on the same side of the imaginary energy axis. For $V_O^->0$ the poles are on the lower half plane, while they move to the upper half plane if $V_O^-<0$. Now, this is the structure of retarded and advanced propagators. Indeed, we recall that if we perturb an action by an external spacetime-dependent source, $\Delta S= \int d^Dx \,J(x) \phi(x)$, the linear response is determined by the retarded Green's function
\be
D_R(x-y)= \Theta(x^0-y^0) \langle[ \phi(x),\phi(y) ]\rangle \,.
\ee
This is manifestly causal: the perturbation at $(y^0, \vec y)$ only affects the dynamics at later times $x^0>y^0$. Consider as an example a free scalar field of mass $m$. In momentum space, the retarded Green's function is given by (see e.g. \cite{Peskin})
\be
D_R(x-y)= \int \frac{dp_0 \, d^{D-1} p}{(2\pi)^D} \,\frac{i}{p_0^2-\vec p^2-m^2+i \epsilon p_0}\,e^{-i p^0(x^0-y^0)+i \vec p \cdot (\vec x- \vec y)}\;,\hspace{10 mm}\epsilon>0\,.
\ee
Comparing with (\ref{eq:poles}), we see that $V_O^->0$ corresponds to a retarded propagator (in lightcone time), and will give a causal result; $V_O^-<0$ gives rise to the advanced propagator with the usual acausality.\footnote{Similarly, we expect that a timelike linear dilaton will give retarded/advanced propagators with respect to $x^0$. But in this case there would be additional physical effects from $D > D_{\text{crit}}$ that need to be taken into account.}

We are now in a position to use the dilaton as a tracer of some of the dynamics contained in this amplitude, following the discussion in \cite{sst, Smatrixpaper}.  Working with kinematic invariants large compared to string scale and using $\Gamma(z)\Gamma(1-z)=\pi/\sin\pi z$ to transform all the $\Gamma$ functions to have positive arguments, we obtain
\bea\label{Aopenfourtransf}
\mathcal A& \sim & \frac{g_{O,0}^{N-2}}{(\sigma^-)^{1/2}} \delta(\sum_I \hat p_I^+)\delta^{D-2}(\sum_I \hat p_{I\perp})\exp\left(-(\sum_{I=1}^N\hat p^-_I-(N-2)i V_O^-)^2/2{\sigma^-}^2\right)\nonumber\\
& &\times \frac{\sin( 2\pi{\ap}(k_A\cdot k_{\hat A}+ k_A\cdot k_B))}{\sin(2\pi {\ap}( k_A\cdot k_{B}-i\epsilon))}\frac{\Gamma(2{\ap} k_A\cdot k_{\hat A}+1)\Gamma(-2{\ap}k_A\cdot k_{\hat A}-2\ap k_A\cdot k_B-1)}{\Gamma(-2{\ap} k_A\cdot k_{B})} \,.
\eea
The $\sin$ factor in the denominator contains the $s$-channel poles.  We can write it in terms of its phases
\be\label{sinexpand}
\sin( 2\pi\ap k_A\cdot k_B)=\frac{1}{2i}\left(e^{-4\pi i\alpha' E^2}e^{2\sqrt{2}\pi\alpha' E V^-_O}-e^{4\pi i\alpha' E^2}e^{-2\sqrt{2}\pi \alpha'E V^-_O}  \right)\,.
\ee
Since $V^-_0>0$, we can expand this in powers of $e^{8\pi i\alpha' E^2}e^{-4\sqrt{2}\pi \alpha' E V^-_O}$, i.e. writing the amplitude as
\bea\label{series}
\mathcal A& \sim & \frac{g_{O,0}^{N-2}}{(\sigma^-)^{1/2}}\delta(\sum_I \hat p_I^+)\delta^{D-2}(\sum_I \hat p_{I\perp})\exp\left(-(\sum_{I=1}^N\hat p^-_I-(N-2)i V_O^-)^2/2{\sigma^-}^2\right)\nonumber\\
& &\times {\sin (2\pi{\ap}(k_A\cdot k_{\hat A}+ k_A\cdot k_B))}\frac{\Gamma(2{\ap} k_A\cdot k_{\hat A}+1)\Gamma(-2{\ap}k_A\cdot k_{\hat A}-2\ap k_A\cdot k_B-1)}{\Gamma(-2{\ap} k_A\cdot k_{B})}\nonumber\\
& &\times e^{4\pi i\alpha' E^2}e^{-2\sqrt{2}\pi \alpha' E V^-_O }\sum_{n=0}^\infty e^{8\pi i n\alpha' E^2}e^{-4\sqrt{2}\pi n \alpha'E V^-_O} \,.
\eea
As explained in section 2.3 of \cite{Smatrixpaper}, the phase in the $n$th term in this expansion corresponds precisely to the time delay $\Delta T$ for the $s$-channel string formed from the joining of $A$ and $B$ to oscillate $n$ times before splitting into $\hat A$ and $\hat B$.  The dilaton dependence fits with that:  the interval in $X^+$ between these events is $\Delta T/\sqrt{2}$, so the series exhibits a factor of $g_O(X^+_{\text{splitting}})$, the string coupling at the $X^+$ position of the splitting interaction.    Similar results hold for closed strings.  

One interesting feature of the above calculation is the way that the dilaton background shifts the Mandelstam invariants so that it plays a role similar to the $i\epsilon$ prescription. In the next subsection, we describe how this comes about in the target QFT. Let us now return to the question of the sign of $V$ in this analysis.  As mentioned above, with the contour oriented as in \cite{Edepsilon}, corresponding to forward time evolution, the integral over the vertex operator positions converges only for $V^->0$ (in physical kinematics with $E>0$).  For this sign of $V^-$, the string coupling becomes weaker at later times, and the infinite series of oscillations (\ref{series}) is well controlled. 

If we wish to consider a solution with the string coupling getting stronger at late times, i.e. $V^-<0$, the integral over vertex operators converges if we reverse the direction of time evolution.  This corresonds to a coupling that weakens in the reversed time direction.  This enables us to consider an infinite series of oscillations in that direction.   (This does not correspond to time advances in the process described in the original forward time direction, it just describes a sequence of processes with the strings joining earlier and earlier, but for each such joining they split later.)  

Altogether, in this basic process we find that the linear dilaton background serves as a reliable tracer of the location of the string interactions.  In this section, we applied it to straightforward string oscillations as a test of the method.  It is interesting to apply this to help trace out the level of non-locality in string interactions \cite{sixpoints}.  

\subsection{QFT reduction}

In this subsection, we reduce this to tree level quantum field theory and analyze $2\to 2$ scattering in a theory with action
\bea\label{QFTaccubic}
S &=&-\int d^D x\, e^{2 V^- x^+}\left(\frac{1}{2}(\partial\eta)^2+\frac{\lambda}{3!}\eta^3 \right) \nonumber\\
&=&- \int d^Dx\, \left(\frac{1}{2} (\partial \eta_c)^2+\frac{\lambda}{3!} e^{-V^- x^+} \eta_c^3\right) \,,
\eea
up to a boundary term, where the canonical field is 
\be\label{etac}
\eta_c=e^{V^- x^+}\eta\,.
\ee
Let us consider $s$-channel tree level $2\to 2$ scattering $AB\to \hat A\hat B$ in this theory.  We introduce a wavepacket for particle $A$ as above in (\ref{wavepacketAGauss}) and convolve that with
\be\label{firstformQFT}
\int d^Dx \, d^Dy\,  \frac{d^Dq}{(2\pi)^D} \, \frac{ie^{-V^-(x^++y^+)}e^{-i q\cdot(x-y)}e^{i(\hat p_A+\hat p_B)\cdot x} e^{i (\hat p_{\hat A}+\hat p_{\hat B})\cdot y}}{q^2-i \epsilon}\,.
\ee
This gives
\be\label{amputated}
{\cal A} =-i\lambda^2(2\pi)^{D-1}\delta(\sum {\hat p}^+)\delta^{D-2}(\sum {\hat p}^\perp) 
\int dx^+ \, dy^+ \,\frac{dq^-}{2\pi} \frac{e^{-x^{+2} \sigma^{-2}/2-V^-(x^++y^+)}e^{i (x^+-y^+)\cdot(q^--{\hat p}_{A0}^--{\hat p}_B^-)}}{-2 ({\hat p}^+_A+{\hat p}^+_B) q^- -i\epsilon}\,.
\ee
Here we made some simplifying choices.  We work in a simple frame with ${\hat p}^\perp_A+{\hat p}^\perp_B=0$, and we choose to consider external momenta satisfying ${\hat p}_{A0}^-+{\hat p}_{B}^-+{\hat p}_{\hat A}^-+{\hat p}_{\hat B}^-=0$.  The latter is not forced on us since translation invariance is broken in the $x^+$ direction, but it is a useful choice for our purpose of tracking the underlying scattering process with the dilaton.   

Next we evaluate the $q^-$ integral.  This has a pole in the lower half plane; the integral gives zero for $x^+>y^+$.  For $y^+>x^+$ we get the residue of the pole,
\be\label{qmint}
{\cal A} =-i\lambda^2(2\pi)^{D-1}\delta^{\perp,+}(\sum {\hat p}) 
\int_{-\infty}^\infty dx^+ \int_{-\infty} ^{x^+}dy^+ (2\pi i) e^{-x^{+2} \sigma^{-2}/2}\frac{e^{-V^-(x^++y^+)}e^{-i (x^+-y^+)\cdot({\hat p}_{A0}^-+{\hat p}_B^-)}}{-2 ({\hat p}^+_A+{\hat p}^+_B)}
\ee    
Let us change variables to $w^+=x^+-y^+$ and integrate first over $x^+$.  This gives
\be\label{propform}
-\lambda^2 (2\pi)^D\delta^{\perp,+}(\sum\hat p)e^{2 V^{-2}/2\sigma^{-2}}\frac{\int_{-\infty}^0 dw^+ e^{V^-w^+} e^{-i w^+(\hat p_{A0}^-+\hat p_B^-)}}{2(\hat p_{A0}^++\hat p_B^+)}
\ee
which converges for $V^->0$, giving a propagator 
\be\label{propQFT}
\sim \frac{1}{(\hat p^+_{A0}+\hat p^+_B)(\hat p_{A0}^-+\hat p^-_B+i V^-)}
\ee
similarly to the string theory behavior we found above.      

\section{Dilaton tracer as a test of string-theoretic nonlocality}

We can use this technique to help trace the interactions in more subtle situations.  In particular, light cone gauge calculations \cite{lennyspreading}\ suggest long range non-locality in the direction of relative motion of strings at large energies compared to string scale.   We reviewed this and generalized it to the linear dilaton background in \S\ref{sec:lc}\ above.     

In this section, we add a linear dilaton to the six point S-matrix amplitude considered in \cite{sixpoints}, a process which exhibits the predicted longitudinal spreading scale as a function of the separation between the central positions of a string and putative detector of its spreading, at the peak of their wavepackets.  As discussed in \cite{sixpoints}, this interaction is parametrically
stronger than appropriate tree-level quantum field theory models with the same kinematics and wavepackets.  Still, the assessment of whether the interaction may occur on the tail of the wavepackets can be subtle, and it is useful to probe that in a different way.  We can use a linear dilaton background to test the predicted long range interaction.      

Let us briefly summarize the system under study.   We have a $3\to 3$ process with strings $A, B, C\to \hat A, \hat B, \hat C$.  We consider kinematics for which $A$ bends slightly into $\hat A$ and so on.  Including a wavepacket for string $A$ as above, we obtain momenta in the nonzero-mode path integral 
\begin{align}\label{ksix}
k_A&=(E_A+q_A^2/(2E_A),-E_A,q_A)+3 i V_O \nonumber\\
k_B&=(E_B+q_B^2/(2E_B),E_B,q_B)-i V_O\nonumber\\
k_C&=(E_C+q_C^2/(2E_C),-E_C,q_C)-i V_O\nonumber\\
{ k}_{\hat A}&=(-E_{\hat A}-q_{\hat A}^2/(2E_{\hat A}),E_{\hat A},q_{\hat A})-iV_O\nonumber\\
{ k}_{\hat B}&=(-E_{\hat B}-q_{\hat B}^2/(2E_{\hat B}),-E_{\hat B},q_{\hat B})-iV_O\nonumber\\
{ k}_{\hat C}&=(-E_{\hat C}-q_{\hat C}^2/(2E_{\hat C}),E_{\hat C},q_{\hat C})-iV_O \nonumber\\
k_7 &= 2 i V_O
\end{align}
where $|q_I|\ll E_I$.  

As discussed in \cite{sixpoints}, the putative detector of string $C$'s spreading is a `detector' string $D=C + \hat C+\hat B$ created in a four-point process $A+B\to \hat A+D$.   String $D$ can interact with string $C$ via a subsequent four-point process $C+D\to \hat C+\hat B$.  This process may occur locally, near the centers of strings $D$ and $C$.  There may also be support for it to occur via the longitudinal spreading of string $C$, detected by string $D$ as predicted in the light cone gauge calculations.  The latter process involves the center of $C$ moving approximately along the trajectory $X^+\sim - E\alpha'$, having localized string $A$ to propagate along $X^+=0$ and produce $D$ at its collision with string $B$ at the origin. For simplicity we take all the energies to be of order $E$.\\ 
\indent The momentum-space amplitude derived in an appropriate regime in \cite{sixpoints}\ (in the absence of a linear dilaton background) takes the simple form 
\be\label{Afourform}
A^{(1)}_4 A^{(2)}_4 B(\alpha' k_D^2, \eta),
\ee
where $B$ is the Beta function. Here $\eta=2\ap( k_B\cdot k_{\hat B}-k_C\cdot k_{\hat C})$ is taken to be non-integer, with the first two factors being the amplitudes for the four-point processes described above.  
Key features of the dynamics follow from the dependence of this amplitude on the kinematic variable
\be\label{kDsq}
k_D^2 \equiv 2(k_C\cdot k_{\hat C}+k_C\cdot k_{\hat B}+k_{\hat B}\cdot k_{\hat C})
\simeq \hat k_D^2 +4 i V_O^-\hat p_{\hat B}^+ = 4 E_{\hat B}(E_C-E_{\hat C})-4\sqrt{2} i V_O^- E_{\hat B} +Q_D^2
\ee
where we have generalized to the linear dilaton theory, and $Q_D^2$ is a quantity of order $\pm {\cal O}(q^2)$ and does not vary strongly with $E_C$.  The variable $E_C$ is conjugate to the separation between the centers of strings $A$ and $C$, and its coupling to $E_{\hat B}\simeq E_D$ in (\ref{kDsq}) is what leads to the appearance of the predicted longitudinal spreading scale in \cite{sixpoints}.   

Other kinematic variables are also shifted by the dilaton as in the open string version of (\ref{KIJs}).  The shift of $k_C\cdot k_{\hat C}$, $k_B\cdot k_{\hat B}$, and $k_A\cdot k_{\hat A}$ are much smaller than the shift of $k_D^2$ in our kinematic regime.  There is a shift of $k_{\hat B}\cdot k_{\hat C}$ which is of the same order ($\sim V^- E$) as the shift of $k_D^2$ (\ref{kDsq}).   This variable enters into one of the auxiliary four point functions, playing a role similar to that in the previous section.   Our interest in the present section has to do with the interactions involving the putative detector string $D$, so we will focus on the effects of the shift in $k_D^2$.     

Before testing the non-local interactions, let us start with a regime containing string oscillations similar to those described above at four points.       For $\alpha'k_D^2\ll \eta <0$, we can write the Beta function in (\ref{Afourform}) as proportional to
\bea\label{kDosc}
B(\alpha'k_D^2, \eta) &\simeq & \frac{\pi \sin(\pi(\alpha'k_D^2+\eta))}{\sin\pi\eta\sin(\pi (\alpha'k_D^2-i\epsilon))}\frac{\Gamma(-\alpha'k_D^2-\eta)}{\Gamma(-\alpha'k_D^2)\Gamma(-\eta)} \nonumber\\
&=& \frac{2\pi i \sin(\pi(\alpha'k_D^2+\eta))}{\sin\pi\eta ~ e^{i\pi (\alpha'k_D^2-i\epsilon)}}\frac{\Gamma(-\alpha'k_D^2-\eta)}{\Gamma(-\alpha'k_D^2)\Gamma(-\eta)}\sum_{n=0}^\infty e^{-2\pi i (\alpha'k_D^2-i\epsilon) n}
\eea              
with all arguments of the $\Gamma$ functions positive (and taken large here for simplicity).    The expansion here is in powers of
\be\label{kDsqexp}
e^{-2\pi i(\alpha' k_D^2-i\epsilon)}\simeq e^{-8\pi i\alpha' (E_{\hat B}(E_C-E_{\hat C})+ Q_D^2/4)} e^{-8\sqrt{2}\pi\alpha' E_{\hat B} V_O^-}
\ee
This has a consistent interpretation, including the power of $g_s$, as a process in which $D$ oscillates $n$ times before interacting with string $C$.  Each oscillation introduces a factor of $g_O(\Delta X^+)^2$.  We have two powers of the open string coupling $g_O$ here, as opposed to one as we had at four points.  That is because we have a four point interaction in this case, as opposed to the 3 point splitting interaction in the four point amplitude.  Having produced $D$ at a finite time, these oscillations extend into the increasingly weakly coupled regime for higher $n$ for $V^->0$.  

Finally, let us work in the regime in which candidate non-local effects arose in \cite{sixpoints}.  We can consider, for example, a Gaussian wavepacket for $E_C$ which has support within $0< \alpha' k_D^2 < -\eta$, for which the Beta function is most simply written as
\be\label{Betasin}
 \frac{\sin(\pi(\alpha' k_D^2+\eta) )\Gamma(\alpha' k_D^2)\Gamma(-\alpha' k_D^2-\eta)}{\Gamma(-\eta)\sin\pi\eta}
\ee
Centering the wavepacket at the minimum of the $\Gamma$ function factors $\alpha'{k_{D}^2}_0=-\eta/2$, as described in 
\cite{sixpoints}\ we can choose parameters such that the momentum space wavefunction times the amplitude has a global maximum there.    This requires a momentum space width (for $\hat p_C$) of order $\sqrt{-\eta}/E\alpha'$, corresponding to a width in position space of order $E\alpha'/\sqrt{-\eta} \ll E\alpha'$.    

We now include the $V$ dependent shift in $k_D^2$ in the same way as before, keeping $V^- E \lesssim 1$ to avoid changing the longitudinal spreading prediction, as described above in \S\ref{sec:lc}.  This also consistently keeps the $\Gamma$ function factors near their extremum, with the dilaton gradient a small perturbation of order $\sim (V^- E)^2/\eta$.       

It is useful to expand the sinusoidal factor in the numerator into its two terms.   One of these generates an apparent early interaction \cite{sixpoints}\ (with $C$ propagating approximately along the trajectory $X^+_C=X^+_A-2\sqrt{2}\pi E_{\hat B}\ap$), according to the peak of the position space wavefunction for the relative position of $C$ and $A$.  The other term generates an apparent delayed interaction by the same amount.   

As we have just seen in studying the regime with oscillations of $D$, the powers of the dilaton precisely track the shifts in $X^+$, with both arising automatically from the factor $e^{i\pi\alpha' k_D^2}$.    
This indicates that the amplitude has an interaction that is displaced from the origin by $\Delta X^+\sim 2\sqrt{2}\pi E_{\hat B}\ap$.          Specifically, we find a single factor of the open string coupling $g_O$ at $\Delta X^+$.  
The relevant interaction according to the longitudinal spreading prediction is the three point interaction from the splitting of $C$ into $C+\hat C$, which may join $D$ near the origin.

\section{Summary and Discussion}

In this work we set up and applied S-matrix perturbation theory in a locally linear dilaton background, focusing on a lightlike variation of the string coupling $g_s(X^+)$.  We worked directly in the physical domain, with real values for the dilaton variation and the momenta of canonically normalized fields and strings.  We employed simple wave packets and worked with a solution $g_s(X^+)$ that keeps interactions everywhere weak.  This tames the worldsheet $X^+$ zero mode integral and enables us to focus on scattering in the linear dilaton region $g_s = g_0 e^{V\cdot X}$.  The effect on the remaining amplitude involves an interesting imaginary shift of the Mandelstam variables, leading to results exhibiting a formal analogy with retarded propagators.  

At weak but nontrivial dilaton variation $V$, we applied this to obtain information on the location of string interactions in scattering processes.  To calibrate our measurement, we first exhibited two examples, at four and six points, with interactions proceeding after any number of oscillations over a scale $\Delta X^+\sim E\alpha'$ greater than the width of our wavepackets.  The dilaton introduces factors of $\exp(-V^-\Delta X^+)$ which precisely correspond to the scale of the oscillations.  We then applied the same technique to putative non-local (but causal) interactions contained in the six point amplitude in a different regime of parameters.  This provides a further test for the predicted scale of interactions mediated by longitudinal spreading described in \cite{sixpoints}: the factors of $g_s$ precisely correspond to interaction at the predicted range.

It will be interesting to extend this to loop diagrams, where the structure of the propagators will have interesting effects.
Another application and motivation for this work arises from the relative genericity of varying dilaton solutions in string theory.  It is important to understand the behavior of string theory at generic dimensionality $D$.  Classically, in the perturbative regime the system consists of a rolling scalar field generating scale factor expansion $a(t)\propto t$  
(in the canonically normalized variables), i.e. marginally sub-accelerated expansion.  This system has a large number of species, including $N_f\sim 2^D$ flavors of axions.   This raises the question of the effect of loop corrections on the dynamics, including the renormalized Newton constant and potential energy relevant for cosmological evolution \cite{SCloops}.  These effects may significantly affect the behavior of generic perturbative string backgrounds, and we hope that the techniques developed here will aid in their evaluation.       

\section*{Acknowledgements}

We would like to thank X. Dong, S. Giddings, B. Kang, S. Shenker, and D. Stanford for various useful discussions of perturbation theory in a linear dilaton background.
The work of E.S.~was supported  in part by the National Science Foundation
under grant PHY-0756174. G.T. is supported by CONICET PIP grant 11220110100533 and by ANPCYT PICT grant 2015-1224.

\newpage

\appendix

\section{Conformal transformations in the linear dilaton CFT}

In this Appendix we discuss the effects of conformal transformations in the linear dilaton CFT, and obtain the change of the momentum $p_\mu$ between the complex plane and the cylinder.

The nontensor terms in (\ref{eq:OPE}) imply that the linear dilaton background changes the conformal transformations of the fields. Recalling that  under an infinitesimal conformal transformation $z'= z+\epsilon v(z)$ the change in a field is given by
\be
\delta A(z_0, \bar z_0) = -\epsilon\, \text{Res}_{z\to z_0}\,v(z)T(z) A(z_0)+\text{h.c.}\,,
\ee
the infinitesimal transformation of $X^\mu$ in a linear dilaton background becomes
\be
\delta X^\mu =-\epsilon \left(v\, \partial X^\mu+\frac{\ap}{2} \partial v\, V^\mu+\ldots \right)+\text{h.c.}
\ee
Similarly,
\be\label{eq:dX1}
\delta (\partial X^\mu)=-\epsilon \left(\partial v \,\frac{\partial X^\mu}{z^2}+\frac{\ap}{2} \partial^2 v \,V^\mu+\ldots \right)\,.
\ee

Next, for a finite transformation $z'=z'(z)$, we propose
\be
(\partial_z z') \partial' X_\mu'(z')= \partial X_\mu(z)+F(z',z) V_\mu
\ee
where the factor on the left hand side is chosen to reproduce the weight $h=1$ part of the conformal transformation. The correct composition with another transformation $z''(z')$ is obtained provided that $F(z'',z)=F(z',z)+\partial_z z' F(z'',z')$. Requiring moreover that $F$ reduces to the previous infinitesimal form, we arrive at
\be
(\partial_z z') \partial' X_\mu'(z')= \partial X_\mu(z)-\frac{\ap}{2}\frac{\partial_z^2 z'}{\partial_z z'} V_\mu\,.
\ee

Let us apply this result to find the conformal map to the cylinder, $w= i \log z$, with $w=\sigma_1+i \sigma_2$,  $\sigma_1 \sim \sigma_1+2\pi\;,\;-\infty<\sigma_2<\infty$. The result is
\be\label{eq:dwX}
\partial_w X^\mu(w) =-i z \partial_z X^\mu(z)-i\frac{\ap}{2} V^\mu\,.
\ee
Finally, let us work out the effects on the creation and annihilation modes. The mode expansion on the complex plane is the same as in the theory with $V_\mu=0$,
\be\label{eq:modes1}
\partial X^\mu(z) =-i \left(\frac{\ap}{2} \right)^{1/2}\sum_{m=-\infty}^\infty\,\frac{\alpha_m^\mu}{z^{m+1}}\;,\;\bar \partial X^\mu(\bar z) =-i \left(\frac{\ap}{2} \right)^{1/2}\sum_{m=-\infty}^\infty\,\frac{\t \alpha_m^\mu}{\bar z^{m+1}}\,,
\ee
with momentum
\be
p^\mu=(2/\ap)^{1/2}\alpha_0^\mu=(2/\ap)^{1/2}\t \alpha_0^\mu\,.
\ee
Mapping to the cylinder and using (\ref{eq:dwX}) gives
\be
\partial_w X^\mu(w)=-\frac{\ap}{2} (p^\mu+i V^\mu)-\left(\frac{\ap}{2} \right)^{1/2} \sum_{m\neq 0} \alpha_m^\mu e^{i m w}\,,
\ee
and similarly for the right-movers. Thus, the change from the plane to the cylinder redefines the momentum by
\be\label{eq:pcyl}
\hat p^\mu=p^\mu+i V^\mu\,.
\ee

\section{Green's functions in linear dilaton field theory}

It is interesting to consider Green's functions in  simple field theory analogues of the linear dilaton background.  Consider as in (\ref{targetEFT}) a QFT with scalar field $\varphi(x)$ and action
\be\label{targetEFTJ}
S= -\int d^D x\,  \sqrt{-g}\, e^{-2\Phi}\left(\frac{1}{2}(\partial \varphi)^2 +J\varphi\right)\,.
\ee
where we have added a source term.  Having packaged the source this way, we have the equation of motion
\be\label{EFTeq}
e^{2\Phi}\partial_\mu (e^{-2\Phi} \partial^\mu \varphi) =J.
\ee
We can solve this via
\be\label{pGJ}
\varphi(x) = \int d^D x' \,G(x, x') J(x')
\ee
with
\be\label{Gvp}
e^{2\Phi}\partial_\mu \left(e^{-2\Phi}\partial^\mu G(x, x')\right) = \delta^{D}(x-x')
\ee
This Green's function can be written as
\be\label{GpV}
G(x, x')=\int \frac{d^D\hat p}{(2\pi)^D}\, \frac{e^{i \hat p\cdot(x-x')}}{\hat p^2+2 i V\cdot\hat p -i \epsilon}
\ee
in the linear dilaton background.  

\begingroup\raggedright\begin{thebibliography}{10}
\baselineskip=14.5pt

\bibitem{sixpoints}

M. Dodelson and E. Silverstein,  ``Long-Range Nonlocality in Six-Point String Scattering:  simulation of black hole infallers'', arxiv:1703.10147 [hep-th].
  
\bibitem{SCloops}

X. Dong, B. Kang, E. Silverstein, G. Torroba, work in progress.  

\bibitem{LDprevious}

C.~T.~Chan and W.~M.~Chen,
  ``Vertex Operators and Scattering Amplitudes of the Bosonic Open String Theory in the Linear Dilaton Background,''
  arXiv:0907.5472 [hep-th].
  
\bibitem{otherLDscatt}  
  
   O.~Aharony, A.~Giveon and D.~Kutasov,
  ``LSZ in LST,''
  Nucl.\ Phys.\ B {\bf 691}, 3 (2004)
  doi:10.1016/j.nuclphysb.2004.05.015
  [hep-th/0404016].

\bibitem{JoeBook}

J.~Polchinski,
  ``String theory. Vol. 1: An introduction to the bosonic string,''
  Cambridge, UK: Univ. Pr. (1998) 402 p

\bibitem{lennyspreading}
L. Susskind,
``Strings, black holes and Lorentz contraction,"
Phys. Rev. D {\bf 49}, 6606-6611 (1994). \\ 
M. Karliner, I. R. Klebanov, and L. Susskind,
``Size and Shape of Strings," 
Int. J. Mod. Phys. {\bf A3} 1981 (1988). 
[hep-th/9308139].

\bibitem{grossmende}
 D.~J.~Gross and P.~F.~Mende,
  ``String Theory Beyond the Planck Scale,''
  Nucl.\ Phys.\ B {\bf 303}, 407 (1988).

\bibitem{BHpaper}
  M.~Dodelson and E.~Silverstein,
  ``String-theoretic breakdown of effective field theory near black hole horizons,''
  arXiv:1504.05536 [hep-th].

\bibitem{Edepsilon} 
  E.~Witten,
  ``The Feynman $i \epsilon$ in String Theory,''
  JHEP {\bf 1504}, 055 (2015)
  [arXiv:1307.5124 [hep-th]].

\bibitem{sst}
N. Seiberg, L. Susskind, and N. Toumbas,
``Space/Time Non-Commutativity and Causality,"
JHEP {\bf{0006}} 044 (2000)  [hep-th/0005015].

\bibitem{Smatrixpaper}
M.~Dodelson and E.~Silverstein,
  ``Longitudinal nonlocality in the string S-matrix,''
  arXiv:1504.05537 [hep-th].

\bibitem{Peskin} 
  M.~E.~Peskin and D.~V.~Schroeder,
  ``An Introduction to quantum field theory,''
  Reading, USA: Addison-Wesley (1995) 842 p


\endgroup
\end{document}